\documentclass[preprint,authoryear,12pt]{elsarticle}
\usepackage{graphicx}
\usepackage{a4wide}
\usepackage{epstopdf}
\usepackage{booktabs}
\usepackage{multirow}
\usepackage{amssymb}
 \usepackage{amsthm}
 \usepackage{amsmath}

\journal{Energy Economics}

\begin{document}

\begin{frontmatter}

\title{Leverage effect in energy futures}

\author[utia,ies]{Ladislav Kristoufek} \ead{kristouf@utia.cas.cz}

\address[utia]{Institute of Information Theory and Automation, Academy of Sciences of the Czech Republic, Pod Vodarenskou Vezi 4, 182 08, Prague, Czech Republic, EU} 
\address[ies]{Institute of Economic Studies, Faculty of Social Sciences, Charles University in Prague, Opletalova 26, 110 00, Prague, Czech Republic, EU}

\begin{abstract}
We propose a comprehensive treatment of the leverage effect, i.e. the relationship between returns and volatility of a specific asset, focusing on energy commodities futures, namely Brent and WTI crude oils, natural gas and heating oil. After estimating the volatility process without assuming any specific form of its behavior, we find the volatility to be long-term dependent with the Hurst exponent on a verge of stationarity and non-stationarity. Bypassing this using by using the detrended cross-correlation and the detrending moving-average cross-correlation coefficients, we find the standard leverage effect for both crude oil. For heating oil, the effect is not statistically significant, and for natural gas, we find the inverse leverage effect. Finally, we also show that none of the effects between returns and volatility is detected as the long-term cross-correlated one. These findings can be further utilized to enhance forecasting models and mainly in the risk management and portfolio diversification.
\end{abstract}


\end{frontmatter}

\newpage

\section{Introduction}

The leverage effect is one of the well-established phenomena of the financial economics. Historically, \cite{Black1976} discusses a possible relationship between returns and changes in volatility of stocks. The argumentation is based on changes in earnings, where decreasing expected earnings of the company push the price down and in turn it decreases the market value of the company which drives the leverage (ratio between debt and equity) up. Negative relationship between returns and volatility is thus referred to as `the leverage effect'. However, in the modern, high-speed, markets where the market prices of assets are driven by many more forces than simple expected earnings, such an explanation of the effect serves as just a little more than an anecdote. The leverage effect can be simply understood as a negative relationship between returns and volatility which are driven by opposite forces. When negative news reaches the market, volatility of the corresponding asset usually increases because of an uncertain future development. Contrarily, the negative news drives the prices down forming a negative return. The leverage effect thus seems as a natural connection of the two characteristics (returns and volatility) of the traded assets.

The leverage effect is usually tightly connected, and sometimes even interchanged, with a notion of the asymmetric volatility. The standard asymmetric volatility is characterized by a lower volatility connected to a bull (growing) market and a higher volatility connected to a bear (declining) market. The definition and interconnection between the two effects -- the leverage effect and the asymmetric volatility -- is thus very close and sometimes hard to distinguish between. Nonetheless, most authors agree on several characteristics of the relationship between returns and volatility -- returns and volatility are negatively correlated, the correlation is quite weak yet still persists over quite long time (with slowly decaying cross-correlations), and the causality goes from returns to volatility and not vice versa \citep{Pagan1996,Bouchaud2001,Bouchaud2001a,Bollerslev2006}.

Here we analyze the leverage effect in the future contracts of energy commodities, namely WTI and Brent crude oils, natural gas and heating oil. We try to provide a coherent treatment of the leverage effect starting from the long-term memory characteristics of volatility and its potential non-stationarity, then moving to the estimation of the correlation between returns and volatility under borderline (non-)stationary and a typical seasonality of futures contracts, and finally checking the slow decay of the cross-correlation function characteristic for long-range cross-correlated processes. We find that the leverage effect in its purest form (significant negative correlation between returns and volatility) is found for two out of four studied commodities. However, the level of correlation is very low -- lower than levels standardly reported for stocks and stock indices. Moreover, we show that the cross-correlations are not identified as hyperbolically decaying, i.e. there are no long-range cross-correlations between returns and volatility of the studied commodities. An important aspect of our analysis stems in not assuming anything about the relationship between returns and volatility which distinguishes our study from the other studies which are majorly built around assuming some kind of asymmetric volatility model, i.e. the leverage effect and asymmetric volatility are assumed ex ante to be frequently found ex post.

The paper is structured as follows. In Section 2, we provide a literature review of recent studies on the leverage effect and asymmetric volatility on energy markets. Section 3 introduces the most important methodological aspects of our work -- volatility estimation, long-term memory and its tests and estimators, estimation of correlations under borderline (non-)stationarity and seasonality, and long-range cross-correlations testing. Section 4 presents the analyzed dataset and results. Section 5 concludes. 

\section{Literature review} 

In this section, we review recent literature on the topic of leverage effect and asymmetric volatility in energy commodities in chronological order.

\cite{Fan2008} examine WTI and Brent crude oil prices with various specifications of the generalized autoregressive conditional heteroskedasticity (GARCH) models for purposes of risk management. They find significant two-way spillover effect between both crude oil markets as well as asymmetric leverage effect in the WTI returns but not in the Brent returns. Interestingly, the uncovered leverage effect implies that positive shocks have much higher impact on the future dynamics of the series than the negative ones which is opposite to the leverage effect found in stocks and it can be thus treated as an inverse leverage effect.

\cite{Zhang2008} study an interrelation between the US dollar exchange rates and crude oil prices with a special focus on spillover effects which they separate into three -- mean spillover, volatility spillover and risk spillover. Apart from a significant long-term cointegration relationship, the authors find significant volatility asymmetry. In a similar way to the previous reference, they find the inverse leverage effect which they attribute mainly the non-renewable property of oil and very different roles and behavior of suppliers and demanders of the commodity. 

\cite{Aloui2009} examine the relationship between crude oil and stock markets utilizing a two regime Markov switching exponential GARCH model. They show that the volatility clustering and the leverage effect can be significantly reduced by allowing for the regime switching. Transition between regimes is mainly connected to economic recessions together with stock markets behavior. \cite{Agnolucci2009} compares predictive powers of GARCH-type and implied volatility models on the WTI future contract. Apart from showing that the GARCH-type models outperform the implied volatility models, the author also finds no leverage effect for the WTI contract. \cite{Cheong2009} then focuses on both WTI and Brent crude oil markets and applies GARCH specification. The author finds that the WTI volatility is more persistent than the one of the Brent crude oil. Even though the leverage effect is found for the Brent market and not for the WTI market, the out-of-sample forecasting exercise provides an evidence that a reduced GARCH model with no asymmetric volatility outperforms the others.

\cite{Wei2010} study both the WTI and Brent futures and compare a wide portfolio of GARCH-type models. Focusing on the performance of 1-day, 5-day and 20-day forecasting, they find that no single model is a clear winner in the horse race of testing. However, the authors favor the non-linear specifications of GARCh which can control for long-term memory as well as asymmetry. Similarly to the previous studies, the results on asymmetry are mixed for the two markets. Even though the asymmetry is found for a strong majority of specifications for the Brent market, the WTI shows mixed evidence.

\cite{Chang2010} focus on the relationship between crude oil and biofuels. Specifically, they are interested in the dynamics of volatility (using the exponential GARCH model) conditional on various phases of the market with respect to the crude oil prices. A significant asymmetric volatility reaction is found only for the soybean futures during the high oil prices. Other futures show no significant asymmetry. \cite{Du2011} examine the linkage between the crude oil volatility and agricultural commodity markets using the stochastic volatility approach in the Bayesian framework. The authors show that speculation, scalping and petroleum investors form important aspects of the volatility formation. In the model, they find a weak leverage effect between instantaneous volatility and prices.

\cite{Reboredo2011} inspects the crude oil dependence structure with various copula functions. He shows that the correlation structure is similar during both bear and bull markets and further states that the crude oil market is strongly globalized. For the favorited model of the marginals -- exponential GARCH -- the volatility asymmetry is found for all studied crude oil series. The same methodology is then applied in \cite{Reboredo2012} where the relationship between oil price and exchange rates is examined. In general, the connection between the oil and exchange rate markets is reported to be very weak. The evidence of volatility asymmetry is mixed as well. \cite{Wu2012} propose a copula-based GARCH model and use it to model dependence between crude oil and the US dollar. In their specification, the leverage effect is not significant for either of the studied futures.

\cite{Chang2012} employs a combined regime switching exponential GARCH model with Student-$t$ distributed error terms to model crude oil futures returns. The model is able to capture the main stylized facts of the crude oil futures. Importantly, the model combines both the regime switching and asymmetric volatility to capture nonlinear dependencies between returns, volatility and higher moments. In accordance to other works, no leverage effect is found for the WTI futures. 

\cite{Ji2012} analyze the effect of crude oil volatility spillovers on non-energy commodities. After controlling for exchange rates, the authors utilize a bivariate exponential GARCH model with time-varying correlation structure. They show that the crude oil plays a core role in the commodities structure as its volatility spills over to other, non-energy, markets as well. The strength of these spillovers even increases after the 2008 financial crisis. Volatility asymmetry is studied as a difference in reaction to bad and good news. The authors find the effect to be significant for majority of the studied pairs.

\cite{Nomikos2012} investigate dynamics of eight energy spot markets on NYMEX. The authors combine a mean-reverting and a spike model with GARCH-type time-varying volatility focusing on risk management issues as well as their forecasting performance. The leverage effect is found for WTI, heating oil and heating oil-WTI crack spread, and the inverse leverage effect is uncovered for gasoline, natural gas, propane and gasoline-WTI crack spread.

Copulas are further utilized by \cite{Tong2013} who study tail dependence between crude oil and refined petroleum markets. Positive dependence is found in both tails so that the markets tend to move together in both bear and bull periods. Asymmetry in tail dependence is found between crude and heating oils, and between crude oil and jet fuel. Interestingly, the upper tail dependence is stronger than in the lower tail for the pre-crisis period. The authors report that the leverage effect, which is found in its standard form, is much stronger for the post-crisis period.

\cite{Salisu2013} study the WTI and Brent crude oil with respect to the structural breaks while controlling for potential volatility asymmetry. Persistence as well as asymmetry of volatility is reported even after controlling for two structural breaks (Iraqi/Kuwait conflict and the financial crisis of 2008) identified for both oil markets. The authors stress that neither of the effects should be studied separately and the constructed models should consider each of structural breaks, volatility persistence and asymmetry.

And \cite{Chkili2014} examine crude oil, natural gas, gold and silver markets using various linear and nonlinear GARCH-type specifications. The nonlinear specifications are found to fare better in a sense of in-sample and out-of-sample performance as well as risk management issues under the Basel II regulations. The direction and significance of the leverage effect are found to be strongly dependent on the model choice.

\section{Methodology}

Studying leverage effect stems primarily on the analysis of the relationship between returns and volatility of the series. As such, this is connected with several issues. Firstly, the volatility itself needs to be extracted from the series. Secondly, the volatility is standardly considered as a long-term memory process. Thirdly, not only is the volatility process long-term dependent but it is usually on the edge of stationarity, i.e. its fractional integration parameter $d \approx 0.5$ and it is thus somewhere between a stationary short-term memory process with $d=0$ and a unit root process with $d=1$. In this section, we introduce methodology and instruments to deal with such specifics.

\subsection{Volatility estimation}

In majority of the leverage effect and asymmetric volatility studies covered in the Literature review, the volatility process has been estimated as a part of the complete model under various assumptions and restrictions. In turn, the volatility series and its characteristics are strongly dependent on the model choice and specifications. For our purposes, the leverage effect emerges from the model only if we assume correlation between the returns and volatility processes. However, if the effect is in reality not present, it can simply occur to be significant during the estimation procedure due to the model misspecification. In our study, we bypass this issue by estimating the volatility outside the returns model.

Historically, the volatility and variance series were estimated simply as a squared or absolute returns of the series. In a sense, the GARCH-type models are built in the same logic. However, these simple measures turn out to be very poor estimators of the true volatility \citep{Chou2010}. Range-based estimators of volatility turn out to be much more efficient and precise than the absolute and squared errors and they stay close to the most efficient realized variance family measures\footnote{We do not opt for the realized variance family measures due to their need of high-frequency data. Moreover, our study is a study of the relationship between returns and volatility, not of finding the best measure of volatility. The range-based estimators are in turn a very fitting compromise as these need only daily open, close, high and low prices which are freely available for practically all publicly traded assets.}.

From several possibilities, we select the Garman--Klass estimator (GKE) as a highly efficient estimator of daily variance. The estimator is defined as
\begin{equation}
\label{GK}
\widehat{\sigma^2_{GK,t}}=\frac{(\log(H_t/L_t))^2}{2}-(2\log2-1)(\log(C_t/O_t))^2
\end{equation}
where $H_t$ and $L_t$ are daily highs and lows, respectively, and $C_t$ and $O_t$ are daily closing and opening prices, respectively \citep{Garman1980}. As the estimator does not take the overnight volatility into consideration, we further work with the open-close returns, i.e. $r_t=\log(C_t)-\log(O_t)$.

\subsection{Long-term memory}

Long-term memory (long memory, long-range dependence) is connected to specific features of the series in both time and frequency domains. In the time domain, the long-term memory process has asymptotically power-law decaying autocorrelation function $\rho(k)$ with lag $k$ such that $\rho(k) \propto k^{2H-2}$ for $k \rightarrow +\infty$. In the frequency domain, the long-term memory process has divergent at origin spectrum $f(\lambda)$ with frequency $\lambda$ such that $f(\lambda)\propto \lambda^{1-2H}$ for $\lambda \rightarrow 0+$. In both definitions, the Hurst exponent $H$ plays a crucial role. For stationary series, $H$ is standardly bounded between 0 and 1 so that $0 \le H <1$. No long-term memory is connected to $H=0.5$, positive long-term autocorrelations are found for $H>0.5$ and negative ones for $H<0.5$. The Hurst exponent is connected to the fractional differencing parameter $d$ in a strict way -- $H=d+0.5$ \citep{Beran1994}.

Hurst exponent is crucial for our further analysis. However, before estimating the exponent itself, we need to test the series for actually being long-range dependent. It has been shown that the estimators of Hurst exponent might report values different from 0.5, and thus hinting long-term memory, even if the series are not long-range dependent \citep{Taqqu1995,Taqqu1996,Teverovsky1999,Lennartz2009,Barunik2010,Kristoufek2010a,Kristoufek2012,Zhou2012}. To deal with this matter, we firstly test for the presence of long-range dependence in the series before estimating the Hurst exponent. We opt for two tests -- modified rescaled range and rescaled variance.

The modified rescaled range test \citep{Lo1991} is an adjusted version of the traditional rescaled range test \citep{Hurst1951} controlling for short-term memory of the series. The testing statistic $V$ is defined as 

\begin{equation}
\label{MRS}
V_{T}=\frac{(R/S)_{T}}{\sqrt{T}}
\end{equation}
where the range $R$ is defined as a difference between the maximum and the minimum of the profile (cumulative demeaned original series), $S$ is the standard deviation of the series and $T$ is the time series length. Here $(R/S)_T$ is the rescaled range of the series of length $T$. To control for the potential short-term memory bias (strong short-term memory might be mistaken for the long-term memory), the standard deviation $S$ is used in its heteroskedasticity and autocorrelation consistent (HAC) version. For these purposes, we utilize the following specification which is later used in the bivariate setting and the rescaled covariance test as well:

\begin{equation}
\label{eq:HAC_CC}
\widehat{s_{xy,q}}=\sum_{k=-q}^q{\left(1-\frac{|k|}{q+1}\right)\widehat{\gamma_{xy}}(k)}
\end{equation}
where $\widehat{\gamma_{xy}}(k)$ is a sample cross-covariance at lag $k$, $q$ is a number of lags taken into consideration and the cross-covariances are weighted with the Barlett-kernel weights. For the purposes of the modified rescaled range, we set $S \equiv \widehat{s_{xx,q}}$ as the autocovariance function is symmetric. We follow the suggestion of \cite{Lo1991} and use lag $q$ according to the following formula for the optimal lag:

\begin{equation}
\label{eq2}
q^{\ast}=\left\lfloor\left(\frac{3T}{2}\right)^{\frac{1}{3}}\left(\frac{2\widehat{|\rho(1)|}}{1-\widehat{\rho(1)}^2}\right)^{\frac{2}{3}}\right\rfloor
\end{equation}
where $\widehat{\rho(1)}$ is a sample first order autocorrelation and $\lfloor \rfloor$ is the lower integer operator. Under the null hypothesis of no long-range dependence, the statistic is distributed as
\begin{equation}
F_V(x)=1+2\sum_{k=1}^{\infty}(1-4k^2x^2)e^{-2(kx)^2}.
\end{equation}

As an alternative to the modified rescaled range test, \cite{Giraitis2003} propose the rescaled variance test which simply substitutes the range in Eq. \ref{MRS} by variance of the profile. The testing statistic $M$ is then defined as

\begin{equation*}
M_T=\frac{var(X)}{TS^2},
\end{equation*}
where $X$ is the profile of the original series and the standard deviation $S$ is defined in the same way as for the modified rescaled range test. \cite{Giraitis2003} show that the rescaled variance test has better properties than the modified rescaled range which is further supported by \cite{Lee1996} and \cite{Lee1997}. Under the null hypothesis of no long-term memory, the statistic is distributed as
\begin{equation}
F_M(x)=1+2\sum^\infty_{k=1}{(-1)^ke^{-2k^2\pi^2x}}.
\end{equation}

For the estimation of the Hurst exponent itself, we utilize two frequency domain estimators -- the local Whittle estimator and the GPH estimator. We opt for the frequency domain estimators as these have well defined asymptotic properties and are well suited even for non-stationary or boundary series which turns out to be the case for the analysis we present.


\cite{Robinson1995a} proposes the local Whittle estimator as a semi-parametric maximum likelihood estimator using the likelihood of \cite{Kunsch1987} while focusing only on a part of spectrum near the origin. As an estimator of the spectrum $f(\lambda)$, the periodogram $I(\lambda)$ is utilized. For the time series of length $T$, and setting $m \le T/2$ and $\lambda_j=2\pi j/T$, the Hurst exponent is estimated as
\begin{equation}
\label{eq:LWX}
\widehat{H}=\arg \min R(H),
\end{equation} 
where 
\begin{equation}
\label{eq:LWX_R}
R(H)=\log\left(\frac{1}{m}\sum_{j=1}^m{\lambda_j^{2H-1}I(\lambda_j)}\right)-\frac{2H-1}{m}\sum_{j=1}^m{\log \lambda_j}.
\end{equation}


\cite{Geweke1983} introduce an estimator based on a full functional specification of the underlying process as the fractional Gaussian noise, which is labeled as the GPH estimator after the authors. The assumption of the underlying process is connected to a specific spectral density which is in turn utilized in the regression estimation of the following equation:

\begin{equation}
\label{GPH}
\log I(\lambda_j) \propto -(H-0.5)\log [4\sin^2(\lambda_j/2)].
\end{equation}

Both estimators are consistent and asymptotically normal. To avoid bias due to short-term memory, we estimate both the local Whittle and GPH estimators only on parts of the estimated periodogram that are close to the origin (short-term memory is present at high frequencies and thus far from the origin). Specifically, we use $m=T^{0.6}$.

\subsection{Correlation coefficient for non-stationary series}

As the leverage effect can be seen as a correlation between returns and volatility, a need for efficient estimators of correlation between potentially non-stationary series is high. Recently, two methods have been proposed in the literature -- detrended cross-correlation coefficient \citep{Zebende2011} and detrending moving-average cross-correlation coefficient \citep{Kristoufek2014a}.

\cite{Zebende2011} proposes the detrended cross-correlation coefficient as a combination of the detrended cross-correlation analysis (DCCA) \citep{Podobnik2008} and the detrended fluctuation analysis (DFA) \citep{Peng1993,Peng1994,Kantelhardt2002}. The detrended cross-correlation coefficient $\rho_{DCCA}(s)$, which measures the correlation even between non-stationary as well as seasonal series, is defined as
\begin{equation}
\rho_{DCCA}(s)=\frac{F^2_{DCCA}(s)}{F_{DFA,x}(s)F_{DFA,y}(s)},
\label{rho}
\end{equation}
where $F^2_{DCCA}(s)$ is a detrended covariance between profiles of the two series based on a window of size $s$, and $F^2_{DFA,x}$ and $F^2_{DFA,y}$ are detrended variances of profiles of the separate series, respectively, for a window size $s$. For more technical details about the methods, please refer to \cite{Kantelhardt2002}, \cite{Podobnik2008} and \cite{Kristoufek2014}. In words, the method is based on calculating the correlation coefficient between series detrended by a linear trend while the detrending is performed in each window of length $s$.

\cite{Kristoufek2014a} introduces the detrending moving-average cross-correlation coefficient as an alternative to the above mentioned coefficient. The method connects the detrending moving average (DMA) procedure \citep{Vandewalle1998,Alessio2002} and detrending moving-average cross-correlation analysis (DMCA) \citep{Arianos2009,He2011a}. The detrending moving-average cross-correlation coefficient $\rho_{DMCA}(\lambda)$ is defined as

\begin{equation}
\rho_{DMCA}(\lambda)=\frac{F_{DMCA}^2(\lambda)}{F_{x,DMA}(\lambda)F_{y,DMA}(\lambda)},
\end{equation}
where $F^2_{DMCA}(\lambda)$, $F^2_{DMA,x}(\lambda)$ and $F^2_{DMA,y}(\lambda)$ are, similarly to the DCCA-based coefficient, detrended covariance between profiles of the two studied series and detrended variances of the separate series, respectively, with a moving average parameter $\lambda$. Contrary to the previous DCCA-based method, the DMCA variant does not require box-splitting but estimates the correlation from the profile series detrended simply by the moving average of length $\lambda$. \cite{Carbone2003} show that the centered moving average outperforms the backward and forward ones so that we apply the centered one in our analysis. For more detailed description of the procedures, please refer to \cite{Alessio2002}, \cite{Arianos2009} and \cite{Kristoufek2014a}.

\subsection{Rescaled covariance test}

Motivated by the rescaled variance test for the univariate series, \cite{Kristoufek2013} proposes the rescaled covariance test which is able to distinguish between long-term and short-term memory between two series. In a similar way as for the univariate series, the long-term memory can be generalized to the bivariate setting so that the long-range cross-correlated (cross-persistent) processes are characterized by asymptotically power-law decaying cross-correlation function and divergent at origin cross-spectrum. By applying the test to the relationship between returns and volatility, we can comment on possible power-law cross-correlated relationship between the two series which is usually connected to the leverage effect \citep{Cont2001}.

The testing statistic for the rescaled covariance test is defined as
\begin{equation}
\label{eq:RC}
M_{xy,T}(q)=q^{\widehat{H_x}+\widehat{H_y}-1}\frac{\widehat{\text{Cov}}(X_T,Y_T)}{T\widehat{s_{xy,q}}},
\end{equation} 
where $\widehat{s_{xy,q}}$ is the HAC-estimator of the covariance of the studied series defined in Eq. \ref{eq:HAC_CC}, $\widehat{\text{Cov}}(X_T,Y_T)$ is the estimated covariance between profiles of the series, and $\widehat{H_x}$ and $\widehat{H_y}$ are estimated Hurst exponents for the separate processes. For the estimated Hurst exponents, we use the average of the local Whittle and GPH estimators if the process is found to be long-range dependent. Otherwise, we set the corresponding exponent equal to 0.5. Critical and $p$-values for the test are obtained from the moving-block bootstrap methodology. For more details, please refer to \cite{Kristoufek2013}.

\section{Data and results}

\begin{figure}[c]
\center
\begin{tabular}{cc}
\includegraphics[width=3.2in]{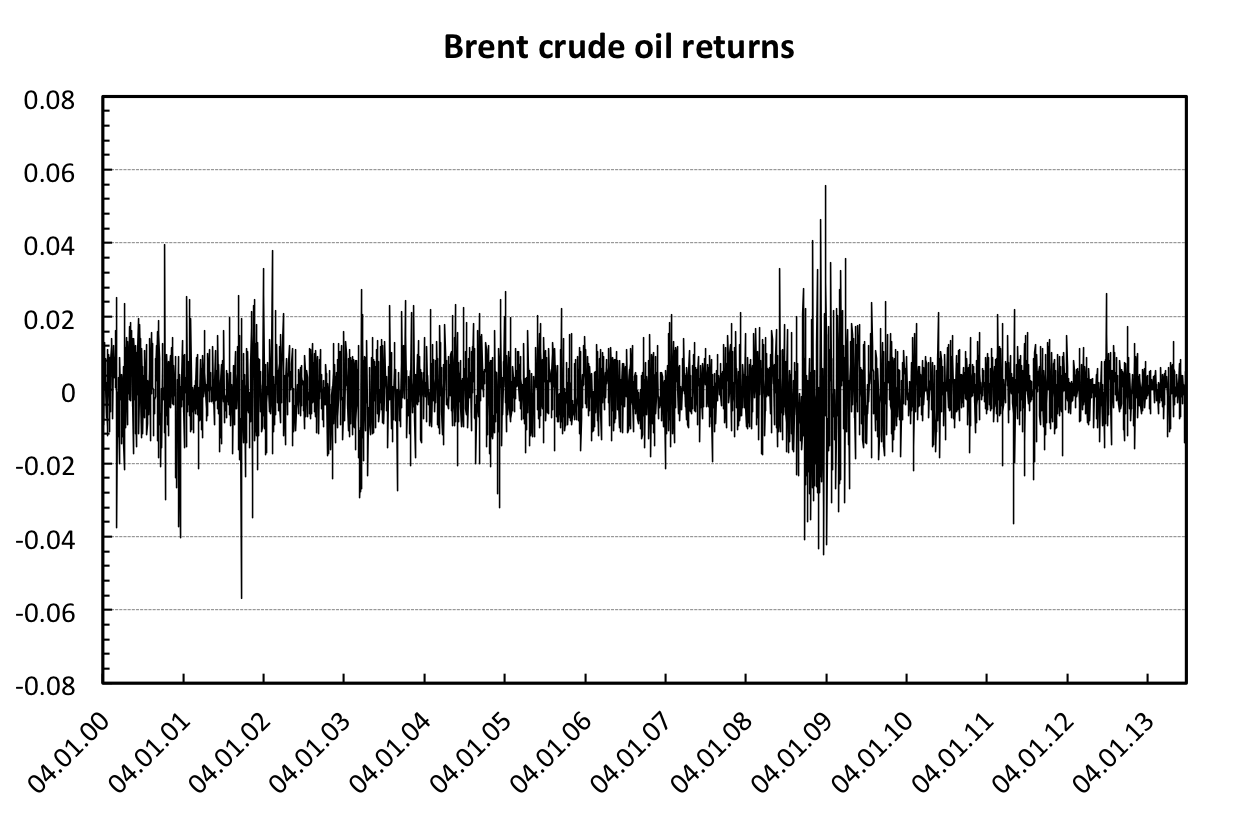}&\includegraphics[width=3.2in]{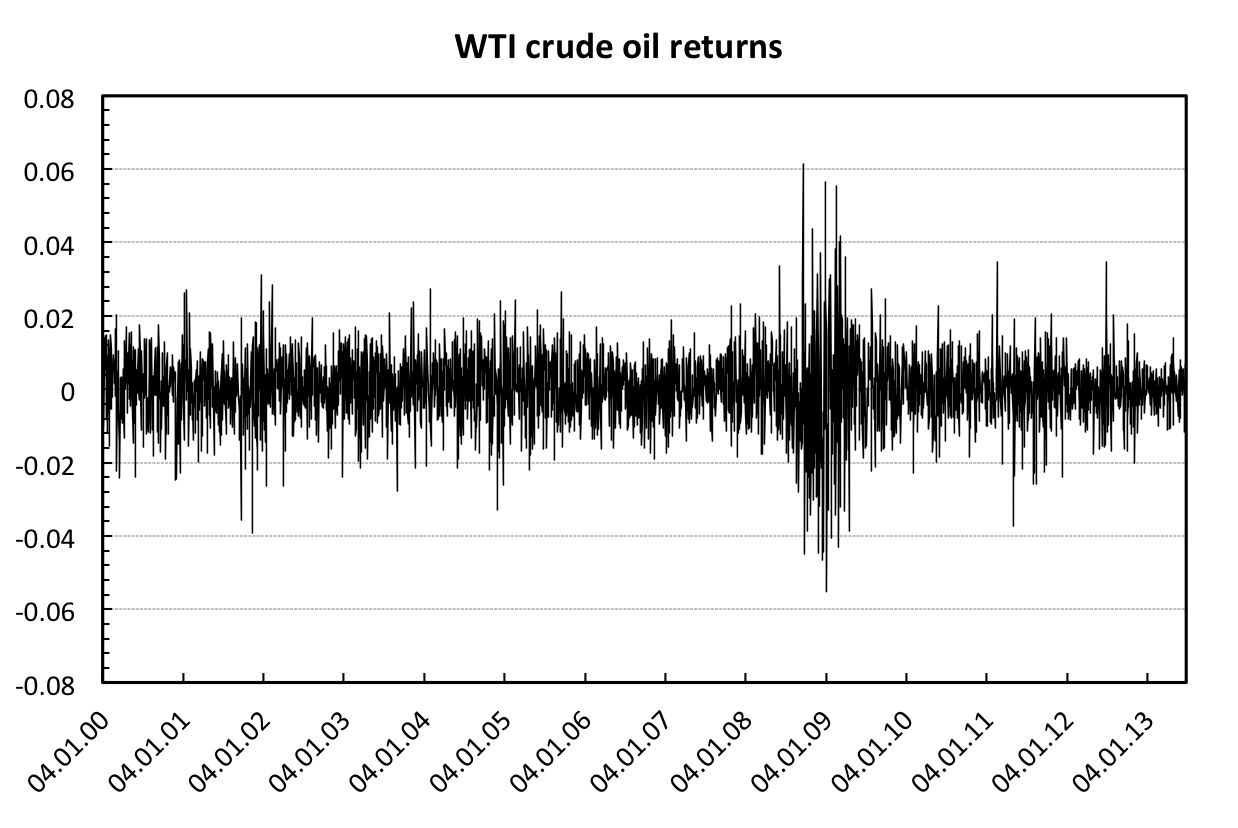}\\
\includegraphics[width=3.2in]{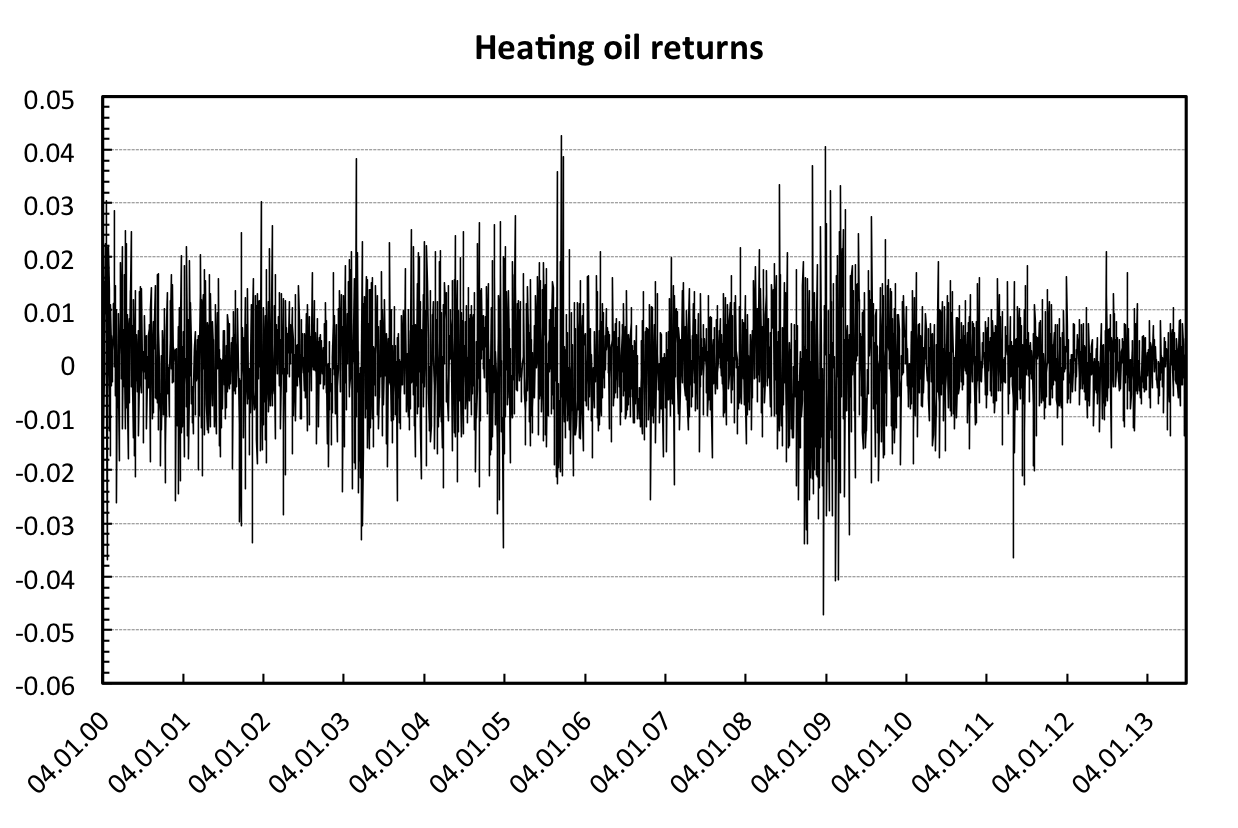}&\includegraphics[width=3.2in]{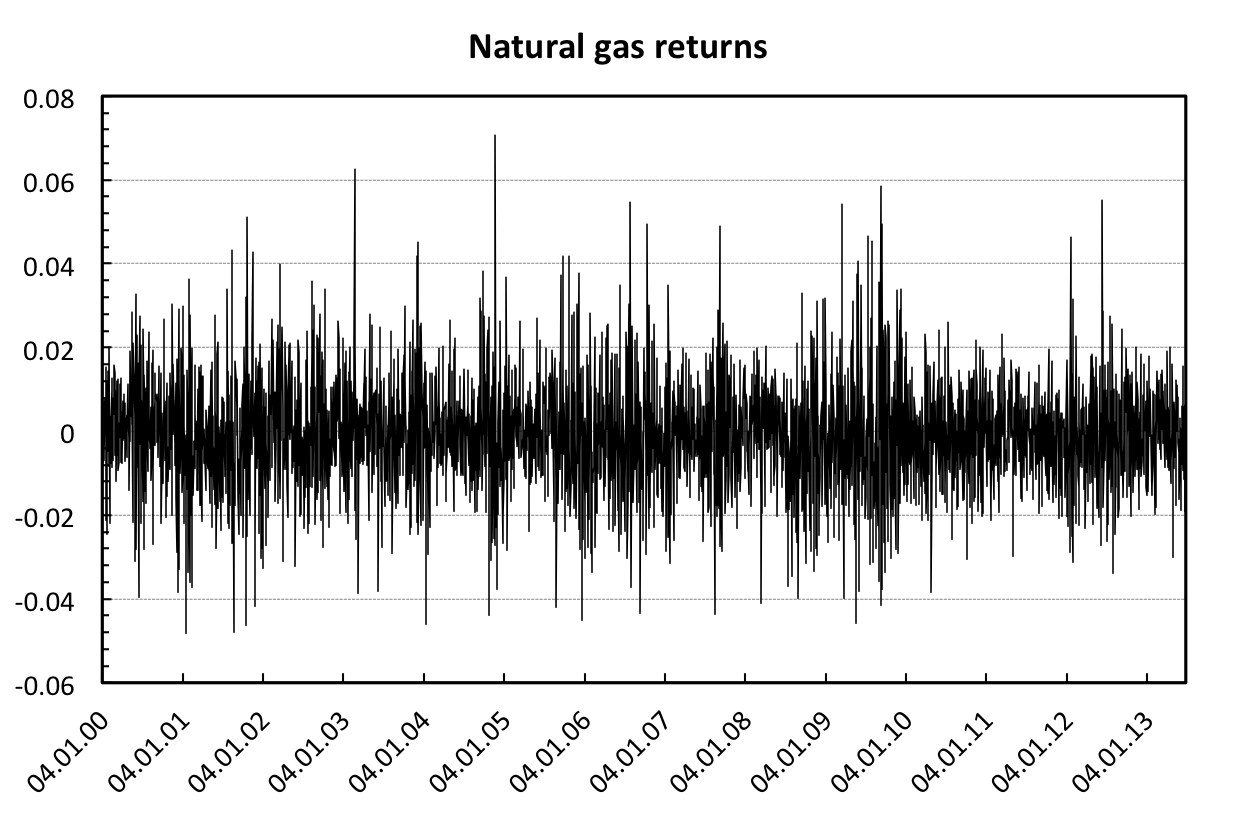}\\
\end{tabular}
\caption{\footnotesize\textbf{Returns of energy futures.} The dynamics of the analyzed futures follow standard patterns of financial returns, mainly non-Gaussian distribution, heavy tails and volatility clustering.\label{fig1}}
\end{figure}

\begin{figure}[c]
\center
\begin{tabular}{cc}
\includegraphics[width=3.2in]{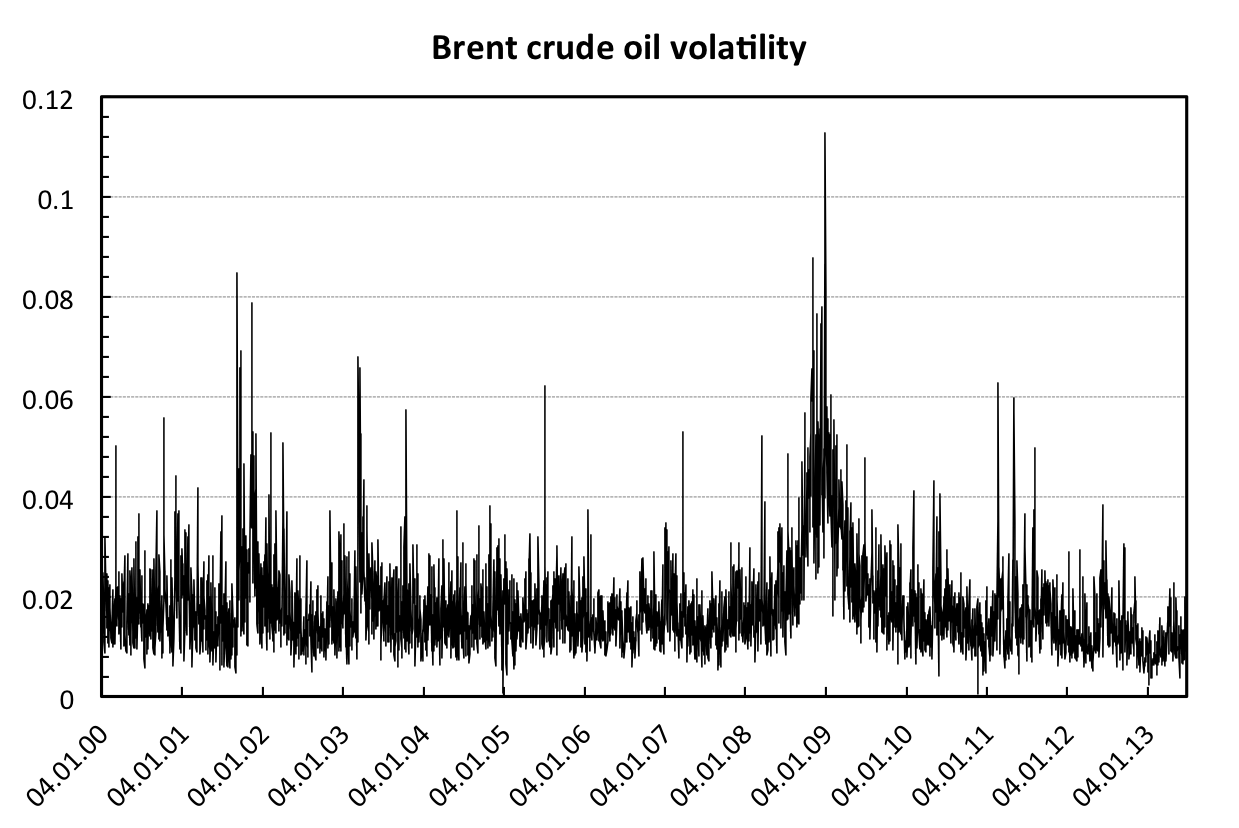}&\includegraphics[width=3.2in]{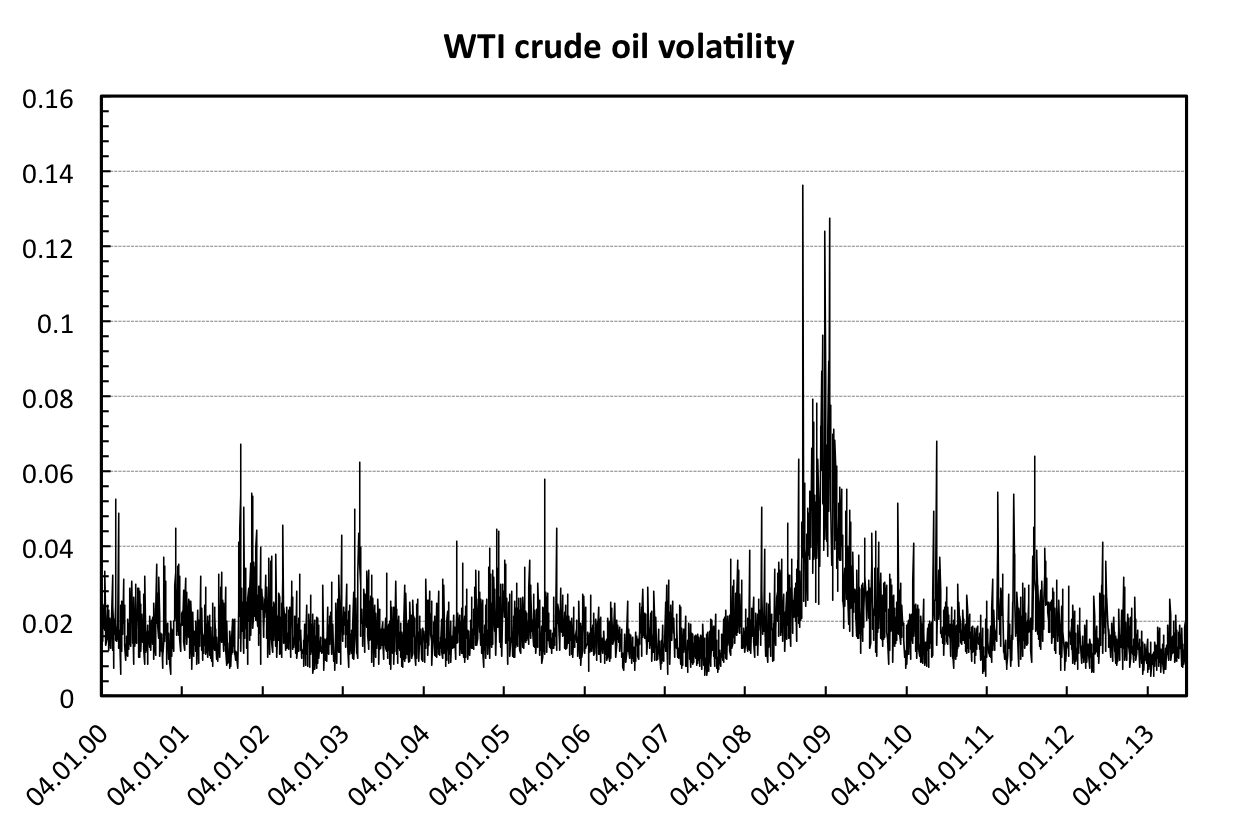}\\
\includegraphics[width=3.2in]{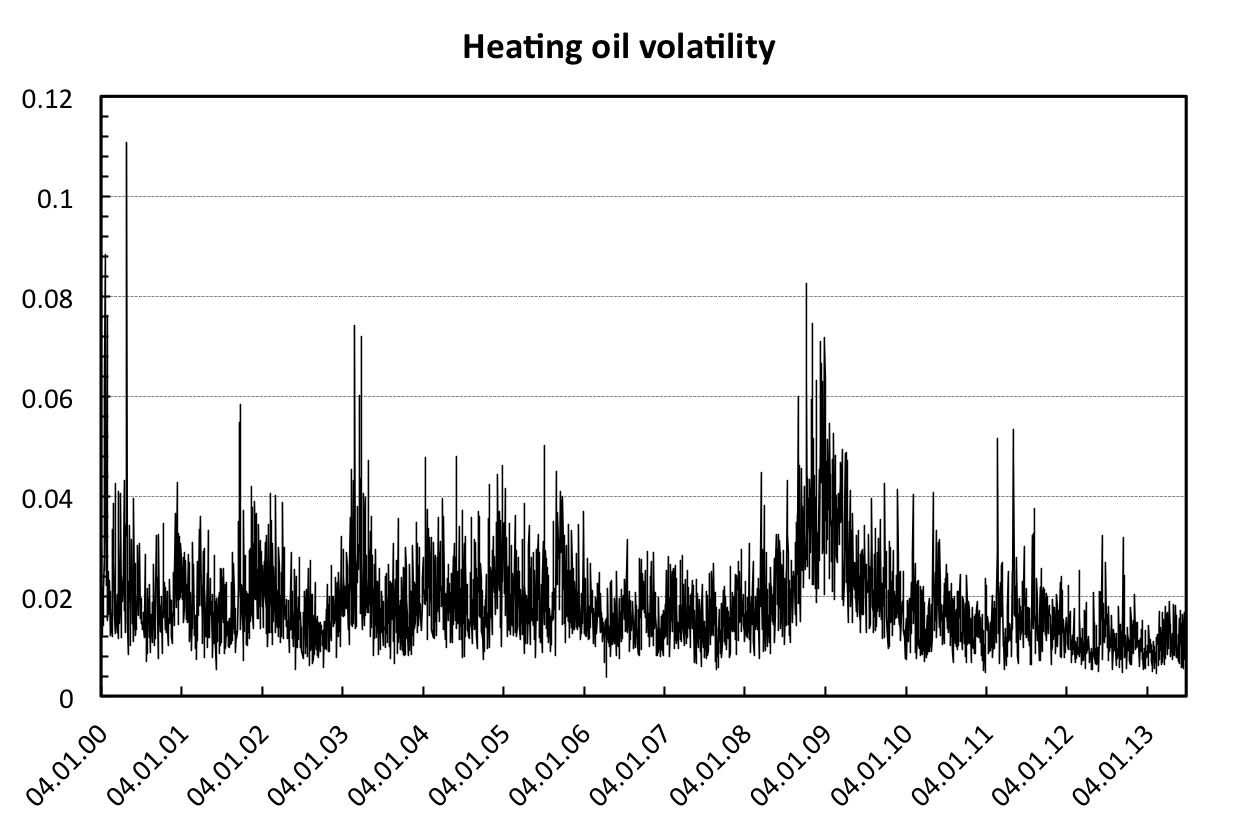}&\includegraphics[width=3.2in]{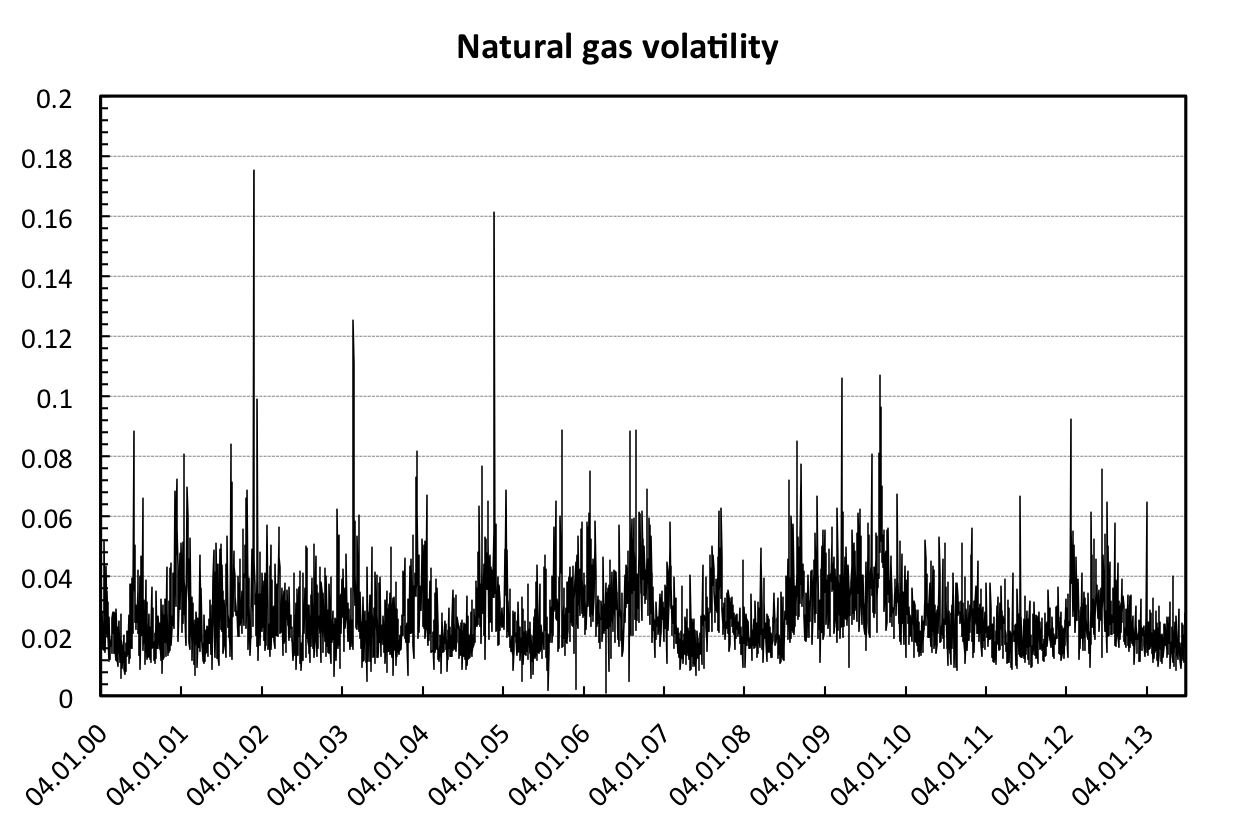}\\
\end{tabular}
\caption{\footnotesize\textbf{Volatility of energy futures.} Volatility follows persistent behavior with strongly varying levels. Again, the dynamics is well in hand with the standard volatility dynamics of financial assets.\label{fig2}}
\end{figure}

We analyze front futures contracts of Brent crude oil, WTI (West Texas Intermediate) crude oil, heating oil and natural gas between 1.1.2000 and 30.6.2013. As we are interested in the leverage effect, we focus on returns and volatility of the future prices. In Figs. \ref{fig1} and \ref{fig2}, we present returns and volatility based on the Garman-Klass estimator given in Eq. \ref{GK}. From the returns charts, we observe that these behave very similarly to the standard financial returns with volatility clustering and non-Gaussian distribution. However, we also notice, mainly for the natural gas, that returns undergo certain seasonal pattern which is connected to the rolling of the front and back futures contracts. This is dealt with by utilizing detrended cross-correlation and detrending moving-average cross-correlation coefficients which are constructed for such seasonalities. Volatility dynamics again reminds of standard volatility of other financial assets with evident persistence, which is dealt with later on. Again, the natural gas series stands out with more frequent volatility jumps and more erratic behavior.

In Tab. \ref{tab1}, we summarize standard descriptive statistics and tests. All returns series follow quite standard characteristics such as excess volatility, negative skewness (apart from natural gas in this case), non-Gaussian distribution and asymptotic stationarity. For each series, we also find significant autocorrelations. Later, we test whether these can be treated as the long-term ones or not. Apart from the returns and volatility, which we examine in its logarithmic form, we focus on the standardized returns as well. Note that the returns standardized by their volatility are usually close to being normally distributed and in general, they are more suitable for statistical analysis. From this point onward, we focus solely on the relationship between standardized returns and logarithmic volatility so that if returns and volatility are referred to, we work with the transformed series. Standardized returns are all approximately symmetric and do not exceed kurtosis of the normal distribution. Moreover, the autocorrelations have been filtered out by standardizing for three out of four series. For the volatility, we strongly reject normality of the distribution and we find very strong autocorrelations. Moreover, we reject both unit root and stationary behavior of the series. This leads us to an inspection of potential long-term memory in the analyzed series.

In Tab. \ref{tab2}, we show results for the modified rescaled range and the rescaled variance tests. Optimal lag has been chosen according to Eq. \ref{eq2}. We find that neither of the returns series are long-range autocorrelated, even though the testing statistics for natural gas are close to the critical levels. As expected, long-term memory is identified for all volatility series even after controlling for rather high number of lags (between 15 and 20). The results of the long-term memory tests thus give expected results -- no long-term memory for the returns and statistically significant long-term memory for the volatility series.

Based on the previous tests, we take that the returns series are not long-term dependent so that their Hurst exponent is equal to 0.5, which is later used in the rescaled covariance test. For the volatility series, we estimate the Hurst exponent $H$ using the local Whittle and GPH estimators. The estimates are summarized in Tab. \ref{tab3}. We observe that both estimators give similar results -- the Hurst exponent for volatility for all four studied series is estimated around $H=1$. Based on the reported standard errors, we cannot distinguish whether the Hurst exponents are below or above the unity value. Therefore, we cannot easily decide whether the volatility series are stationary long-range dependent or non-stationary long-range dependent but still mean reverting. Nevertheless, this does not discredit any of the following instruments and tests.

As the volatility series are long-term correlated, we need to apply correlation measures which are able to deal with such series. \cite{Kristoufek2014} shows that the standard correlation coefficient is not able to do so. We thus apply the detrended cross-correlation coefficient and detrending moving-average cross-correlation measures which are not only able to work under long-term memory and even non-stationarity but they can also filter out well-defined trends. In the case of the studied futures, the rolling period of a trading month is well-established so that we can set $s=\lambda=20$ and the methods filter the seasonality away. Tab. \ref{tab4} reports the estimated correlation coefficients between returns and volatility of each studied commodity. We find that both crude oils are partially driven by the standard leverage effect connected to negative correlation between returns and volatility. For heating oil, the estimated correlation is also negative but not statistically significant\footnote{$p$-values are constructed using 10,000 series generated using Fourier randomization, which ensures that the autocorrelation structure remains untouched but the cross-correlations are shuffled away.} at 1\% level. Natural gas is then characterized by the inverse leverage effect, i.e. the positive correlation between returns and volatility. Note that even though some of the correlations are found to be statistically significant, the levels are rather weak compared to standardly reported ones for stocks or stock indices.

Tab. \ref{tab5} then summarizes the results of the rescaled covariance test which test possible long-range cross-correlations. We use the same number of lags as for the univariate volatility tests in Tab. \ref{tab3}. Based on the reported $p$-values, we find no sings of long-range dependence in the bivariate setting. This is tightly connected to rather weak correlations found above. Even though the series might be correlated, creating the leverage or the inverse leverage effects, the influence is not strong enough to translate into a long-term connection.

\section{Conclusion}

In this paper, we propose a comprehensive treatment of the leverage effect, focusing on energy commodities futures, namely Brent and WTI crude oils, natural gas and heating oil. After estimating the volatility process without assuming any specific form of its behavior, we find the volatility to be long-term dependent with the Hurst exponent on a verge of stationarity and non-stationarity. Bypassing this using by using the detrended cross-correlation and the detrending moving-average cross-correlation coefficients, we find the standard leverage effect for both crude oils. For heating oil, the effect is not statistically significant, and for natural gas, we find the inverse leverage effect. This points out a need for initial testing for the presence of the leverage effect before constructing any specific models to avoid inefficient estimation or even biased results. Finally, we also show that none of the effects between returns and volatility is detected as the long-term cross-correlated one. The dynamics of the crude oil futures, as ones of the most traded ones, is thus closer to the one of stocks and stock indices whereas the less popular heating oil and natural gas somewhat deviate from the standard behavior. These findings can be further utilized to enhance forecasting models and mainly in the risk management and portfolio diversification.

\section*{Acknowledgements}
The research leading to these results has received funding from the European Union's Seventh Framework Programme (FP7/2007-2013) under grant agreement No. FP7-SSH-612955 (FinMaP). Support from the Czech Science Foundation under projects No. P402/11/0948 and No. 14-11402P is also gratefully acknowledged.

\section*{References}

\bibliographystyle{chicago}
\bibliography{Leverage_EE}

\begin{thebibliography}{}

\bibitem[\protect\citeauthoryear{Agnolucci}{Agnolucci}{2009}]{Agnolucci2009}
Agnolucci, P. (2009).
\newblock Volatility in crude oil futures: A comparison of the predictive
  ability of {GARCH} and implied volatility models.
\newblock {\em Energy Economics\/}~{\em 31}, 316--321.

\bibitem[\protect\citeauthoryear{Alessio, Carbone, Castelli, and
  Frappietro}{Alessio et~al.}{2002}]{Alessio2002}
Alessio, E., A.~Carbone, G.~Castelli, and V.~Frappietro (2002).
\newblock Second-order moving average and scaling of stochastic time series.
\newblock {\em European Physical Journal B\/}~{\em 27}, 197--200.

\bibitem[\protect\citeauthoryear{Aloui and Jammazi}{Aloui and
  Jammazi}{2009}]{Aloui2009}
Aloui, C. and R.~Jammazi (2009).
\newblock The effects of crude oil on stock market shifts behaviour: A regime
  switching approach.
\newblock {\em Energy Economics\/}~{\em 31}, 789--799.

\bibitem[\protect\citeauthoryear{Arianos and Carbone}{Arianos and
  Carbone}{2009}]{Arianos2009}
Arianos, S. and A.~Carbone (2009).
\newblock Cross-correlation of long-range correlated series.
\newblock {\em Journal of Statistical Mechanics: Theory and Experiments\/}~{\em
  3}, P03037.

\bibitem[\protect\citeauthoryear{Barunik and Kristoufek}{Barunik and
  Kristoufek}{2010}]{Barunik2010}
Barunik, J. and L.~Kristoufek (2010).
\newblock On {H}urst exponent estimation under heavy-tailed distributions.
\newblock {\em Physica A\/}~{\em 389(18)}, 3844--3855.

\bibitem[\protect\citeauthoryear{Beran}{Beran}{1994}]{Beran1994}
Beran, J. (1994).
\newblock {\em Statistics for Long-Memory Processes}, Volume~61 of {\em
  Monographs on Statistics and Applied Probability}.
\newblock New York: Chapman and Hall.

\bibitem[\protect\citeauthoryear{Black}{Black}{1976}]{Black1976}
Black, F. (1976).
\newblock Studies on stock price volatility changes.
\newblock {\em Proceedings of the 1976 Meetings of the American Statistical
  Association, Business and Economic Statistics\/}, 177--181.

\bibitem[\protect\citeauthoryear{Bollerslev, Litvinova, and Tauchen}{Bollerslev
  et~al.}{2006}]{Bollerslev2006}
Bollerslev, T., J.~Litvinova, and G.~Tauchen (2006).
\newblock Leverage and volatility feedback effects in high-frequency data.
\newblock {\em Journal of Financial Econometrics\/}~{\em 4(3)}, 353--384.

\bibitem[\protect\citeauthoryear{Bouchaud, Matacz, and Potters}{Bouchaud
  et~al.}{2001}]{Bouchaud2001a}
Bouchaud, J.-P., A.~Matacz, and M.~Potters (2001).
\newblock Leverage effect in financial markets: the retarded volatility model.
\newblock {\em Physical Review Letters\/}~{\em 87}, 228701.

\bibitem[\protect\citeauthoryear{Bouchaud and Potters}{Bouchaud and
  Potters}{2001}]{Bouchaud2001}
Bouchaud, J.-P. and M.~Potters (2001).
\newblock More stylized facts of financial markets: leverage effect and
  downside correlations.
\newblock {\em Physica A\/}~{\em 299}, 60--70.

\bibitem[\protect\citeauthoryear{Carbone and Castelli}{Carbone and
  Castelli}{2003}]{Carbone2003}
Carbone, A. and G.~Castelli (2003).
\newblock Scaling properties of long-range correlated noisy signals:
  application to financial markets.
\newblock {\em Proceedings to SPIE\/}~{\em 5114}, 406--414.

\bibitem[\protect\citeauthoryear{Chang}{Chang}{2012}]{Chang2012}
Chang, K.-L. (2012).
\newblock Volatility regimes, asymmetric basis effects and forecasting
  performance: An empirical investigation of the {WTI} crude oil futures
  market.
\newblock {\em Energy Economics\/}~{\em 34}, 294--306.

\bibitem[\protect\citeauthoryear{Chang and Su}{Chang and Su}{2010}]{Chang2010}
Chang, T.-H. and H.-M. Su (2010).
\newblock The substitutive effect of biofuels on fossil fuels in the lower and
  higher crude oil price periods.
\newblock {\em Energy\/}~{\em 35}, 2807--2813.

\bibitem[\protect\citeauthoryear{Cheong}{Cheong}{2009}]{Cheong2009}
Cheong, C.-W. (2009).
\newblock Modeling and forecasting crude oil markets using {ARCH}-type models.
\newblock {\em Energy Policy\/}~{\em 37}, 2346--2355.

\bibitem[\protect\citeauthoryear{Chkili, Hammoudeh, and Nguyen}{Chkili
  et~al.}{2014}]{Chkili2014}
Chkili, W., S.~Hammoudeh, and D.~Nguyen (2014).
\newblock Volatility forecasting and risk management for commodity markets in
  the presence of asymmetry and long memory.
\newblock {\em Energy Economics\/}~{\em 41}, 1--18.

\bibitem[\protect\citeauthoryear{Chou, Chou, and Liu}{Chou
  et~al.}{2010}]{Chou2010}
Chou, R.-Y., H.-C. Chou, and N.~Liu (2010).
\newblock {\em Handbook of Quantitative Finance and Risk Management}, Chapter
  Range Volatility Models and Their Appliations in Finance, pp.\  1273--1281.
\newblock Springer.

\bibitem[\protect\citeauthoryear{Cont}{Cont}{2001}]{Cont2001}
Cont, R. (2001).
\newblock Empirical properties of asset returns: stylized facts and statistical
  issues.
\newblock {\em Quantitative Finance\/}~{\em 1(2)}, 223 -- 236.

\bibitem[\protect\citeauthoryear{Du, Lu, and Hayes}{Du et~al.}{2011}]{Du2011}
Du, X., C.~Lu, and D.~Hayes (2011).
\newblock Speculation and volatility spillover in the crude oil and
  agricultural commodity markets: A {Bayesian} analysis.
\newblock {\em Energy Economics\/}~{\em 33}, 497--503.

\bibitem[\protect\citeauthoryear{Fan, Zhang, Tsai, and Wei}{Fan
  et~al.}{2008}]{Fan2008}
Fan, Y., Y.-J. Zhang, H.-T. Tsai, and Y.-M. Wei (2008).
\newblock Estimating {`Value at Risk'} of crude oil price and its spillover
  effect using the {GED}-{GARCH} approach.
\newblock {\em Energy Economics\/}~{\em 30}, 3156--3171.

\bibitem[\protect\citeauthoryear{Garman and Klass}{Garman and
  Klass}{1980}]{Garman1980}
Garman, M. and M.~Klass (1980).
\newblock On the estimation of security price volatilities from historical
  data.
\newblock {\em Journal of Business\/}~{\em 53}, 67--78.

\bibitem[\protect\citeauthoryear{Geweke and Porter-Hudak}{Geweke and
  Porter-Hudak}{1983}]{Geweke1983}
Geweke, J. and S.~Porter-Hudak (1983).
\newblock The estimation and application of long memory time series models.
\newblock {\em Journal of Time Series Analysis\/}~{\em 4(4)}, 221--238.

\bibitem[\protect\citeauthoryear{Giraitis, Kokoszka, Leipus, and
  Teyssiere}{Giraitis et~al.}{2003}]{Giraitis2003}
Giraitis, L., P.~Kokoszka, R.~Leipus, and G.~Teyssiere (2003).
\newblock Rescaled variance and related tests for long memory in volatility and
  levels.
\newblock {\em Journal of Econometrics\/}~{\em 112}, 265--294.

\bibitem[\protect\citeauthoryear{He and Chen}{He and Chen}{2011}]{He2011a}
He, L.-Y. and S.-P. Chen (2011).
\newblock A new approach to quantify power-law cross-correlation and its
  application to commodity markets.
\newblock {\em Physica A\/}~{\em 390}, 3806--3814.

\bibitem[\protect\citeauthoryear{Hurst}{Hurst}{1951}]{Hurst1951}
Hurst, H. (1951).
\newblock Long term storage capacity of reservoirs.
\newblock {\em Transactions of the American Society of Engineers\/}~{\em 116},
  770--799.

\bibitem[\protect\citeauthoryear{Ji and Fan}{Ji and Fan}{2012}]{Ji2012}
Ji, Q. and Y.~Fan (2012).
\newblock How does oil price volatility affect non-energy commodity markets?
\newblock {\em Applied Energy\/}~{\em 89}, 273--280.

\bibitem[\protect\citeauthoryear{Kantelhardt, Zschiegner, Koscielny-Bunde,
  Bunde, Havlin, and Stanley}{Kantelhardt et~al.}{2002}]{Kantelhardt2002}
Kantelhardt, J., S.~Zschiegner, E.~Koscielny-Bunde, A.~Bunde, S.~Havlin, and
  E.~Stanley (2002).
\newblock {Multifractal Detrended Fluctuation Analysis of Nonstationary Time
  Series}.
\newblock {\em Physica A\/}~{\em 316(1-4)}, 87--114.

\bibitem[\protect\citeauthoryear{Kristoufek}{Kristoufek}{2010}]{Kristoufek2010a}
Kristoufek, L. (2010).
\newblock Rescaled range analysis and detrended fluctuation analysis: {F}inite
  sample properties and confidence intervals.
\newblock {\em {AUCO} {C}zech {E}conomic {R}eview\/}~{\em 4}, 236--250.

\bibitem[\protect\citeauthoryear{Kristoufek}{Kristoufek}{2012}]{Kristoufek2012}
Kristoufek, L. (2012).
\newblock How are rescaled range analyses affected by different memory and
  distributional properties? {A} {M}onte {C}arlo study.
\newblock {\em Physica A\/}~{\em 391}, 4252--4260.

\bibitem[\protect\citeauthoryear{Kristoufek}{Kristoufek}{2013}]{Kristoufek2013}
Kristoufek, L. (2013).
\newblock Testing power-law cross-correlations: Rescaled covariance test.
\newblock {\em European Physical Journal B\/}~{\em 86}, art. 418.

\bibitem[\protect\citeauthoryear{Kristoufek}{Kristoufek}{2014a}]{Kristoufek2014a}
Kristoufek, L. (2014a).
\newblock Detrending moving-average cross-correlation coefficient: Measuring
  cross-correlations between non-stationary series.
\newblock {\em Physica A\/}~{\em submitted}, 0--0.

\bibitem[\protect\citeauthoryear{Kristoufek}{Kristoufek}{2014b}]{Kristoufek2014}
Kristoufek, L. (2014b).
\newblock Measuring correlations between non-stationary series with {DCCA}
  coefficient.
\newblock {\em Physica A\/}~{\em 402}, 291--298.

\bibitem[\protect\citeauthoryear{K\"unsch}{K\"unsch}{1987}]{Kunsch1987}
K\"unsch, H. (1987).
\newblock Statistical aspects of self-similar processes.
\newblock {\em Proceedings of the First World Congress of the Bernoulli
  Society\/}~{\em 1}, 67--74.

\bibitem[\protect\citeauthoryear{Lee and Schmidt}{Lee and
  Schmidt}{1996}]{Lee1996}
Lee, D. and P.~Schmidt (1996).
\newblock On the power of the kpss test of stationarity against
  fractionally-integrated alternatives.
\newblock {\em Journal of Econometrics\/}~{\em 73}, 285--302.

\bibitem[\protect\citeauthoryear{Lee and Amsler}{Lee and
  Amsler}{1997}]{Lee1997}
Lee, H. and C.~Amsler (1997).
\newblock Consistency of the kpss unit root against fractionally integrated
  alternative.
\newblock {\em Economics Letters\/}~{\em 55}, 151--160.

\bibitem[\protect\citeauthoryear{Lennartz and Bunde}{Lennartz and
  Bunde}{2009}]{Lennartz2009}
Lennartz, S. and A.~Bunde (2009).
\newblock Eliminating finite-size effects and detecting the amount of white
  noise in short records with long-term memory.
\newblock {\em Physical Review E\/}~{\em 79}, 066101.

\bibitem[\protect\citeauthoryear{Lo}{Lo}{1991}]{Lo1991}
Lo, A. (1991).
\newblock Long-term memory in stock market prices.
\newblock {\em Econometrica\/}~{\em 59(5)}, 1279--1313.

\bibitem[\protect\citeauthoryear{Nomikos and Adriosopoulos}{Nomikos and
  Adriosopoulos}{2012}]{Nomikos2012}
Nomikos, N. and K.~Adriosopoulos (2012).
\newblock Modelling energy spot prices: Empirical evidence from {NYMEX}.
\newblock {\em Energy Economics\/}~{\em 34}, 1153--1169.

\bibitem[\protect\citeauthoryear{Pagan}{Pagan}{1996}]{Pagan1996}
Pagan, A. (1996).
\newblock The econometrics of financial markets.
\newblock {\em Journal of Empirical Finance\/}~{\em 3}, 15--102.

\bibitem[\protect\citeauthoryear{Peng, Buldyrev, Goldberger, Havlin, Simons,
  and Stanley}{Peng et~al.}{1993}]{Peng1993}
Peng, C., S.~Buldyrev, A.~Goldberger, S.~Havlin, M.~Simons, and H.~Stanley
  (1993).
\newblock Finite-size effects on long-range correlations: Implications for
  analyzing {DNA} sequences.
\newblock {\em Physical Review E\/}~{\em 47(5)}, 3730--3733.

\bibitem[\protect\citeauthoryear{Peng, Buldyrev, Havlin, Simons, Stanley, and
  Goldberger}{Peng et~al.}{1994}]{Peng1994}
Peng, C., S.~Buldyrev, S.~Havlin, M.~Simons, H.~Stanley, and A.~Goldberger
  (1994).
\newblock Mosaic organization of {DNA} nucleotides.
\newblock {\em Physical Review E\/}~{\em 49\/}(2), 1685--1689.

\bibitem[\protect\citeauthoryear{Podobnik and Stanley}{Podobnik and
  Stanley}{2008}]{Podobnik2008}
Podobnik, B. and H.~Stanley (2008).
\newblock Detrended cross-correlation analysis: A new method for analyzing two
  nonstationary time series.
\newblock {\em Physical Review Letters\/}~{\em 100}, 084102.

\bibitem[\protect\citeauthoryear{Reboredo}{Reboredo}{2011}]{Reboredo2011}
Reboredo, J. (2011).
\newblock How do crude oil prices co-move? a copula approach.
\newblock {\em Energy Economics\/}~{\em 33}, 948--955.

\bibitem[\protect\citeauthoryear{Reboredo}{Reboredo}{2012}]{Reboredo2012}
Reboredo, J. (2012).
\newblock Modelling oil price and exchange rate co-movements.
\newblock {\em Journal of Policy Modeling\/}~{\em 34}, 419--440.

\bibitem[\protect\citeauthoryear{Robinson}{Robinson}{1995}]{Robinson1995a}
Robinson, P. (1995).
\newblock Gaussian semiparametric estimation of long range dependence.
\newblock {\em The Annals of Statistics\/}~{\em 23(5)}, 1630--1661.

\bibitem[\protect\citeauthoryear{Salisu and Fasanya}{Salisu and
  Fasanya}{2013}]{Salisu2013}
Salisu, A. and I.~Fasanya (2013).
\newblock Modelling oil price volatility with structural breaks.
\newblock {\em Energy Policy\/}~{\em 52}, 554--562.

\bibitem[\protect\citeauthoryear{Taqqu, Teverosky, and Willinger}{Taqqu
  et~al.}{1995}]{Taqqu1995}
Taqqu, M., W.~Teverosky, and W.~Willinger (1995).
\newblock Estimators for long-range dependence: an empirical study.
\newblock {\em Fractals\/}~{\em 3\/}(4), 785--798.

\bibitem[\protect\citeauthoryear{Taqqu and Teverovsky}{Taqqu and
  Teverovsky}{1996}]{Taqqu1996}
Taqqu, M. and V.~Teverovsky (1996).
\newblock {On Estimating the Intensity of Long-Range Dependence in Finite and
  Infinite Variance Time Series}.
\newblock In {\em A Practical Guide To Heavy Tails: Statistical Techniques and
  Applications}.

\bibitem[\protect\citeauthoryear{Teverovsky, Taqqu, and Willinger}{Teverovsky
  et~al.}{1999}]{Teverovsky1999}
Teverovsky, V., M.~Taqqu, and W.~Willinger (1999).
\newblock A critical look at lo's modified r/s statistic.
\newblock {\em Journal of Statistical Planning and Inference\/}~{\em 80(1-2)},
  211--227.

\bibitem[\protect\citeauthoryear{Tong, Wu, and Zhou}{Tong
  et~al.}{2013}]{Tong2013}
Tong, B., C.~Wu, and C.~Zhou (2013).
\newblock Modeling the co-movements between crude oil and refined petroleum
  markets.
\newblock {\em Energy Economics\/}~{\em 40}, 882--897.

\bibitem[\protect\citeauthoryear{Vandewalle and Ausloos}{Vandewalle and
  Ausloos}{1998}]{Vandewalle1998}
Vandewalle, N. and M.~Ausloos (1998).
\newblock Crossing of two mobile averages: A method for measuring the roughness
  exponent.
\newblock {\em Physical Review E\/}~{\em 58}, 6832--6834.

\bibitem[\protect\citeauthoryear{Wei, Wang, and Huang}{Wei
  et~al.}{2010}]{Wei2010}
Wei, Y., Y.~Wang, and D.~Huang (2010).
\newblock Forecasting crude oil market volatility: Further evidence using
  {GARCH}-class models.
\newblock {\em Energy Economics\/}~{\em 32}, 1477--1484.

\bibitem[\protect\citeauthoryear{Wu, Chung, and Chang}{Wu
  et~al.}{2012}]{Wu2012}
Wu, C.-C., H.~Chung, and Y.-H. Chang (2012).
\newblock The economic value of co-movement between oil price and exchange rate
  using copula-based {GARCH} models.
\newblock {\em Energy Economics\/}~{\em 34}, 270--282.

\bibitem[\protect\citeauthoryear{Zebende}{Zebende}{2011}]{Zebende2011}
Zebende, G. (2011).
\newblock {DCCA} cross-correlation coefficient: Quantifying level of
  cross-correlation.
\newblock {\em Physica A\/}~{\em 390}, 614--618.

\bibitem[\protect\citeauthoryear{Zhang, Fan, Tsai, and Wei}{Zhang
  et~al.}{2008}]{Zhang2008}
Zhang, Y.-J., Y.~Fan, H.-T. Tsai, and Y.-M. Wei (2008).
\newblock Spillover effect of us dollar exchange rate on oil prices.
\newblock {\em Journal of Policy Modeling\/}~{\em 30}, 973--991.

\bibitem[\protect\citeauthoryear{Zhou}{Zhou}{2012}]{Zhou2012}
Zhou, W.-X. (2012).
\newblock Finite-size effect and the components of multifractality in financial
  volatility.
\newblock {\em Chaos, Solitons and Fractals\/}~{\em 45}, 147--155.

\end{thebibliography}

\newpage

\begin{table}[c]
\centering
\caption{Descriptive statistics}
\label{tab1}
\footnotesize
\begin{tabular}{c|c|cccc}
\toprule \toprule
&&Brent Crude oil&WTI Crude oil&Heating Oil&Natural Gas\\
\midrule \midrule
&mean&0.0000&0.0001&0.0001&-0.0006\\
&SD&0.0089&0.0094&0.0092&0.0131\\
&skewness &-0.2045&-0.1524&-0.0618&0.2035\\
&ex. kurtosis&3.0454&3.5389&1.6150&1.6458\\
&Jarque-Bera&1358&1770&368&403\\
raw&$p$-value&$<0.01$&$<0.01$&$<0.01$&$<0.01$\\
returns&Q(30)&56.5784&93.2582&48.8078&55.6547\\
&$p$-value&$<0.01$&$<0.01$&0.0160&$<0.01$\\
&ADF&-9.3723&-8.9838&-9.0643&-12.7284\\
&$p$-value&$<0.01$&$<0.01$&$<0.01$&$<0.01$\\
&KPSS&0.0729&0.2095&0.0936&0.4147\\
&$p$-value&$>0.1$&$>0.1$&$>0.1$&0.0700\\
\midrule
&mean&0.0189&0.0313&0.0004&-0.0217\\
&SD&0.4530&0.4359&0.4569&0.4342\\
&skewness &0.0051&0.0032&0.0046&0.0536\\
&ex. kurtosis&-0.3861&-0.5599&-0.4995&-0.5268\\
&Jarque-Bera&21&44&35&41\\
standardized&$p$-value&$<0.01$&$<0.01$&$<0.01$&$<0.01$\\
returns&Q(30)&38.8799&49.0375&37.5385&38.4963\\
&$p$-value&0.1280&0.0160&0.1620&0.1370\\
&ADF&-13.4725&-9.9805&-8.9793&-12.1317\\
&$p$-value&$<0.01$&$<0.01$&$<0.01$&$<0.01$\\
&KPSS&0.1994&0.1133&0.0604&0.7664\\
&$p$-value&$>0.1$&$>0.1$&$>0.1$&$<0.01$\\
\midrule
&mean&-4.1419&-4.0664&-4.0993&-3.7100\\
&SD&0.4612&0.4348&0.4422&0.4448\\
&skewness &0.0974&0.5432&0.1736&-0.0990\\
&ex. kurtosis&2.0910&0.9238&0.2861&1.4772\\
&Jarque-Bera&634&285&28&311\\
logarithmic&$p$-value&$<0.01$&$<0.01$&$<0.01$&$<0.01$\\
volatility&Q(30)&11663&16121&14720&7948\\
&$p$-value&$<0.01$&$<0.01$&$<0.01$&$<0.01$\\
&ADF&-4.1167&-4.1354&-3.77728&-5.2721\\
&$p$-value&$<0.01$&$<0.01$&$<0.01$&$<0.01$\\
&KPSS&2.0720&1.0965&5.2119&0.7198\\
&$p$-value&$<0.01$&$<0.01$&$<0.01$&0.013\\
\bottomrule \bottomrule
\end{tabular}
\end{table}

\begin{table}[c]
\centering
\caption{Long-term memory tests}
\label{tab2}
\footnotesize
\begin{tabular}{c|c|cccc}
\toprule \toprule
&&Brent Crude oil&WTI Crude oil&Heating Oil&Natural Gas\\
\midrule \midrule
&$V_T$&1.4603&1.5970&0.7816&1.6995\\
raw&$p$-value&$>0.1$&$>0.1$&$>0.1$&0.0564\\
returns&$M_T$&0.0742&0.1141&0.0218&0.1809\\
&$p$-value&$>0.1$&$>0.1$&$>0.1$&0.0551\\
&$q^{\ast}$&2&1&0&4\\
\midrule
&$V_T$&1.5398&1.5729&1.1521&2.0182\\
standardized&$p$-value&$>0.1$&$>0.1$&$>0.1$&0.0171\\
returns&$M_T$&0.1058&0.0970&0.0663&0.2520\\
&$p$-value&$>0.1$&$>0.1$&$>0.1$&0.0281\\
&$q^{\ast}$&2&2&1&2\\
\midrule
&$V_T$&2.6776&2.8426&3.2812&2.6474\\
logarithmic&$p$-value&$<0.01$&$<0.01$&$<0.01$&$<0.01$\\
volatility&$M_T$&0.6971&0.5708&0.8271&0.4882\\
&$p$-value&$<0.01$&$<0.01$&$<0.01$&$<0.01$\\
&$q^{\ast}$&18&20&19&15\\
\bottomrule \bottomrule
\end{tabular}
\end{table}

\begin{table}[c]
\centering
\caption{Estimated Hurst exponents for logarithmic volatility}
\label{tab3}
\footnotesize
\begin{tabular}{c|cccc}
\toprule \toprule
&Brent Crude oil&WTI Crude oil&Heating Oil&Natural Gas\\
\midrule \midrule
Local Whittle&1.0448&1.1008&1.0803&1.0659\\
st. error&0.0437&0.0440&0.0440&0.0440\\
GPH&1.0383&1.0987&1.1861&0.9838\\
st. error&0.0580&0.0612&0.0660&0.0640\\
\midrule
average&1.0415&1.0997&1.0832&1.0248\\
\bottomrule \bottomrule
\end{tabular}
\end{table}

\begin{table}[c]
\centering
\caption{Estimated correlation coefficients between standardized returns and logarithmic volatility}
\label{tab4}
\footnotesize
\begin{tabular}{c|cccc}
\toprule \toprule
&Brent Crude oil&WTI Crude oil&Heating Oil&Natural Gas\\
\midrule \midrule
$\rho_{DCCA}(s)$&-0.1704&-0.1934&-0.0755&0.1110\\
$p$-value&$<0.01$&$<0.01$&0.0599&$<0.01$\\
$\rho_{DMCA}(\lambda)$&-0.0536&-0.0811&-0.0179&0.0518\\
$p$-value&$<0.01$&$<0.01$&0.0240&$<0.01$\\
\bottomrule \bottomrule
\end{tabular}
\end{table}

\begin{table}[c]
\centering
\caption{Rescaled covariance test for standardized returns and logarithmic volatility}
\label{tab5}
\footnotesize
\begin{tabular}{c|cccc}
\toprule \toprule
&Brent Crude oil&WTI Crude oil&Heating Oil&Natural Gas\\
\midrule \midrule
$M_{xy,T}(q)$&-70.4810&68.5844&117.7159&-452.8543\\
$p$-value&$>0.1$&$>0.1$&$>0.1$&$>0.1$\\
\bottomrule \bottomrule
\end{tabular}
\end{table}

\end{document}